\documentclass{article}
\usepackage{arxiv}
\usepackage{indentfirst}
\usepackage[utf8]{inputenc} 
\usepackage[T1]{fontenc}    
\usepackage{hyperref}       
\hypersetup{
    colorlinks=true,
    linkcolor=blue,
    filecolor=magenta,      
    urlcolor=cyan,
}
\usepackage{url}            
\usepackage{booktabs}       
\usepackage{amsfonts}       
\usepackage{nicefrac}       
\usepackage{microtype}      
\usepackage{lipsum}
\usepackage[square,numbers]{natbib}
\usepackage{graphicx}
\usepackage[ruled,vlined]{algorithm2e}
\usepackage{amsmath}
\usepackage{float} 
\usepackage{xcolor,colortbl}
\usepackage{changepage}  

\begin{document}

\renewcommand*{\thefootnote}{\arabic{footnote}}

\title{Three Errors and Two Problems in a Recent Paper:\\
\textit{gazeNet}: End-to-end eye-movement event detection with deep neural networks (Zemblys, Niehorster, and Holmqvist, 2019)}

\author{
  Lee Friedman\\
  Department of Computer Science\\
  Texas State University\\
  San Marcos, Texas, USA, 78666\\
  \texttt{lfriedman10@gmail.com} \\
}

\maketitle

\begin{abstract}
Zemblys et al. \cite{gazeNet} reported on a method for the classification of eye-movements ("gazeNet"). I have found 3 errors and two problems with that paper that are explained herein. \underline{\textit{\textbf{Error 1:}}} The gazeNet classification method was built assuming that a hand-scored dataset from Lund University was all collected at 500 Hz, but in fact, six of the 34 recording files were actually collected at 200Hz.  Of the six datasets that were used as the training set for the gazeNet algorithm, 2 were actually collected at 200Hz. \underline{\textit{\textbf{Problem 1}}} has to do with the fact that even among the 500Hz data, the inter-timestamp intervals varied widely.  \underline{\textit{\textbf{Problem 2}}} is that there are many unusual discontinuities in the saccade trajectories from the Lund University dataset that make it a very poor choice for the construction of an automatic classification method.  \underline{\textit{\textbf{Error 2}}} The gazeNet algorithm was trained on the Lund dataset, and then compared to other methods, not trained on this dataset, in terms of performance on this dataset.  This is an inherently unfair comparison, and yet no where in the gazeNet paper is this unfairness mentioned. \underline{\textit{\textbf{Error 3}}} arises out of the novel event-related agreement analysis employed by the gazeNet authors.  Although the authors intended to classify unmatched events as either false positives or false negatives, many are actually being classified as true negatives.  True negatives are not errors, and any unmatched event  misclassified as a true negative is actually driving kappa  higher, whereas unmatched events should be driving kappa lower.
\end{abstract}

\section{Error Number 1}
\label{ErrorNumber1}
The gazeNet paper depends on the hand scoring of the eye-tracker data from Lund University (hereafter called the \textit{Lund2013}\footnote{Available for download at\\ http://www.humlab.lu.se/en/person/MarcusNystrom/}.)  The first error has to do with this dataset, on which the gazeNet algorithm was trained and developed.  Here is how the gazeNet authors describe the \textit{Lund2013}:
\\

\begin{adjustwidth}{1cm}{}
``It consists of monocular eye-movement data of participants viewing images, videos, and moving dots. The eye-movements of all participants were recorded with the SMI Hi-Speed 1250 eye-tracker, running at a sampling frequency of 500 Hz. Two domain experts then manually segmented data into fixations, saccades, post-saccadic oscillations (PSO), smooth pursuit, blinks and undefined events. A comprehensive description of the \textit{Lund2013} dataset and the coding process can be found in \cite{Larsson}.'' (page 3, left column, last paragraph)
\end{adjustwidth}

\noindent
\\\noindent
\\The 34 recordings where subjects were observing images were included in their analysis. Unfortunately, the gazeNet authors were unaware of the fact that in six\footnote{Higlighted in blue in Appendix 1} 
of the 34 files  the median inter-timestamp intervals were 5000 microseconds, for a nominal sample rate of 200 Hz rather than the expected 500Hz.  Two of these 200 Hz datasets were included in the gazeNet training set (of a total of 6), one each was in the validation sets for rater RA and rater MN, and one each was in the test sets for rater RA and rater MN.  The gazeNet paper states that ``Trials were 4 or 10 s long.'' (column 2, page 3). This statement is false.  If one subtracts the first timestamp from the last timestamp, one will see that all of the files are almost exactly 10 seconds.  Since the gazeNet algorithm was trained and validated on improper data, it would seem to the present author that, at the least, the algorithm is not optimally trained. Since the gazeNet model was trained on improper data, there is basis for concern about the value of the reported gazeNet values in the gazeNet paper.

\section{Problem 1}
The gazeNet authors apply the MNH \cite{MNH} to the \textit{Lund2013}. The MNH \cite{MNH} was developed using the EyeLink 1000, at a 1000Hz sampling rate.  (I created the MNH.)  It is worth considering whether the application of the MNH to the \textit{Lund2013} is appropriate.  Most of the \textit{Lund2013} signals were sampled near 500 Hz, but as noted above, some of the data were sampled at 200 Hz.  Although the gazeNet authors correctly adjusted the sampling frequency parameter of the MNH to 500 Hz prior to application, they obviously did not adjust either the MNH or the data for the recordings at 200 Hz. Furthermore, even for studies sampled at 500Hz, there is significant variability in the inter-timestamp intervals in the \textit{Lund2013} (Figure 1).  According to technical staff at SR-Research, the inter-timestamp intervals for the EyeLink 1000 are all exactly 1 msec.  So, prior to applying the MNH to this data, this instability in the sampling rate would need to be addressed. \footnote{As the creator of the MNH, I cannot endorse the application of the MNH to data collected at 500 Hz.  I do not have extensive experience assessing its performance on such data. It was written for a reading task recorded with the EyeLink 1000 at 1000Hz.  However, I would suggest that those who intend to use it on such data anyway, interpolate their data to 1000 Hz using the “Piecewise Cubic Hermite Interpolating Polynomial” (PCHIP) method available in MATLAB.}

\begin{figure}[htbp]
\center
\frame{\includegraphics[width=0.4\textwidth]{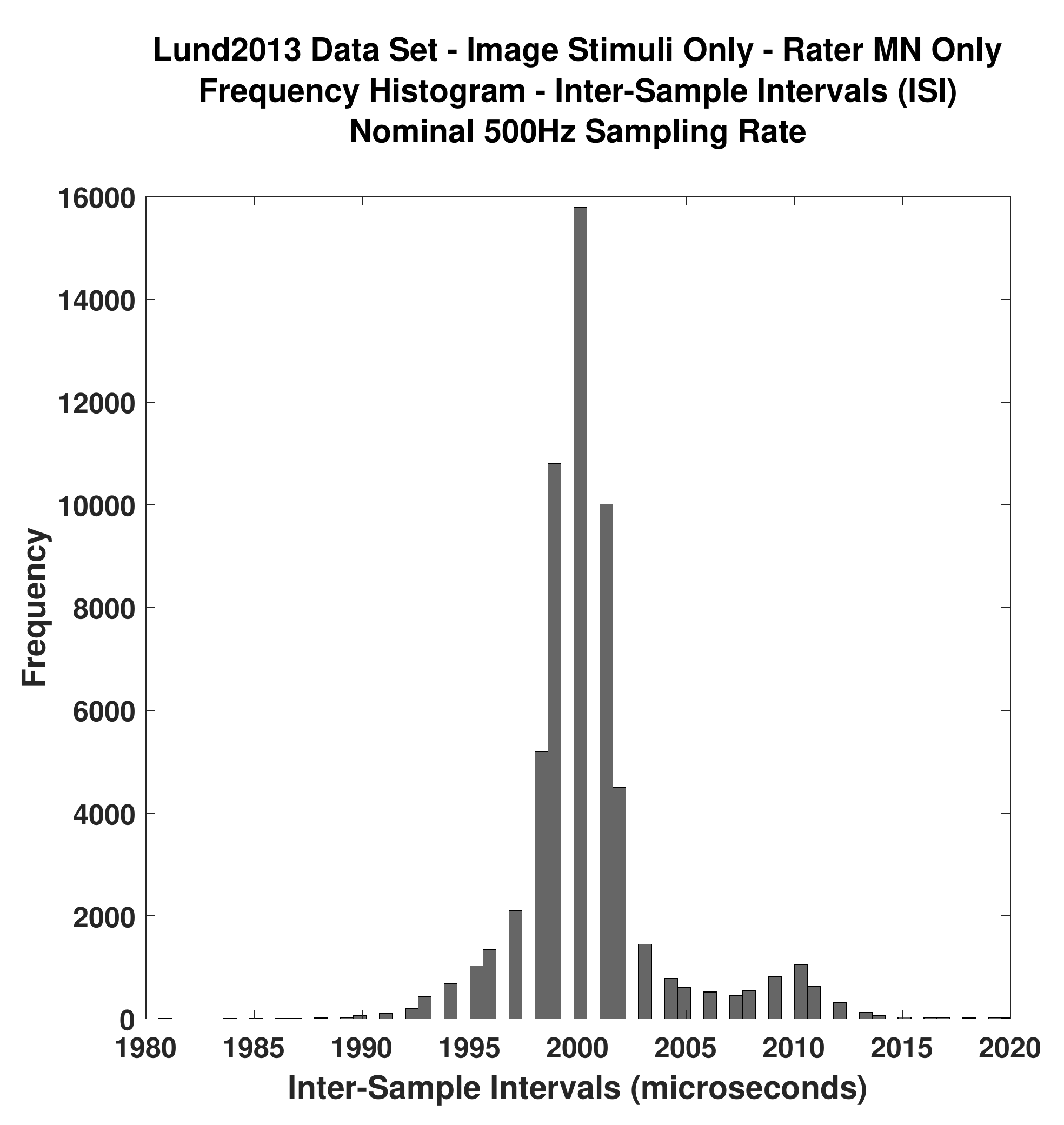}}
\caption{Inter-Sample Intervals (ISI) for the \textit{Lund2013}.  This histogram reflects all the data from all the time stamps for the fourteen 500Hz (nominal) recordings that were coded by rater MN. At that sample rate, samples should be exactly 2000 microseconds apart. Note the bimodal nature of this histogram and the considerable variability around the nominal rate.}
\label{fig:1}  
\end{figure}

\section{Problem 2}
There is a further problem with the \textit{Lund2013}.  A simple inspection of the recordings in the \textit{Lund2013} reveals many discontinuities in saccades. (See Figure 2 for examples.)  In a large number of samples during saccades (for this analysis only files scored by MN were used), there is a large gap in horizontal pixel position followed by a very short gap in pixel position. The blue dots in Figure 2 point to the sample after a large horizontal position move and followed by a small move.  To automatically detect these events, saccade samples were searched to find samples where the horizontal distance from the previous sample was greater than 3 times larger than the distance to the next sample.  Also, the distance to the next sample had to be less than 5.0 pixels.  Since such events can occur naturally at the end of saccades, the search was stopped 6 samples before the end of each saccade. There were 292 saccades in the entire dataset, and 109 (37\%) had a least 1 of these unusual events detected (see Figure 3).  The detection was not perfect, but the vast majority were indeed these discontinuities. Three saccades had 4 such events.  These discontinuities tend to occur more often in large saccades, and also at a time when eye-velocity is high.  

\begin{figure*}[htbp]
\center
\frame{\includegraphics[width=1.0\textwidth]{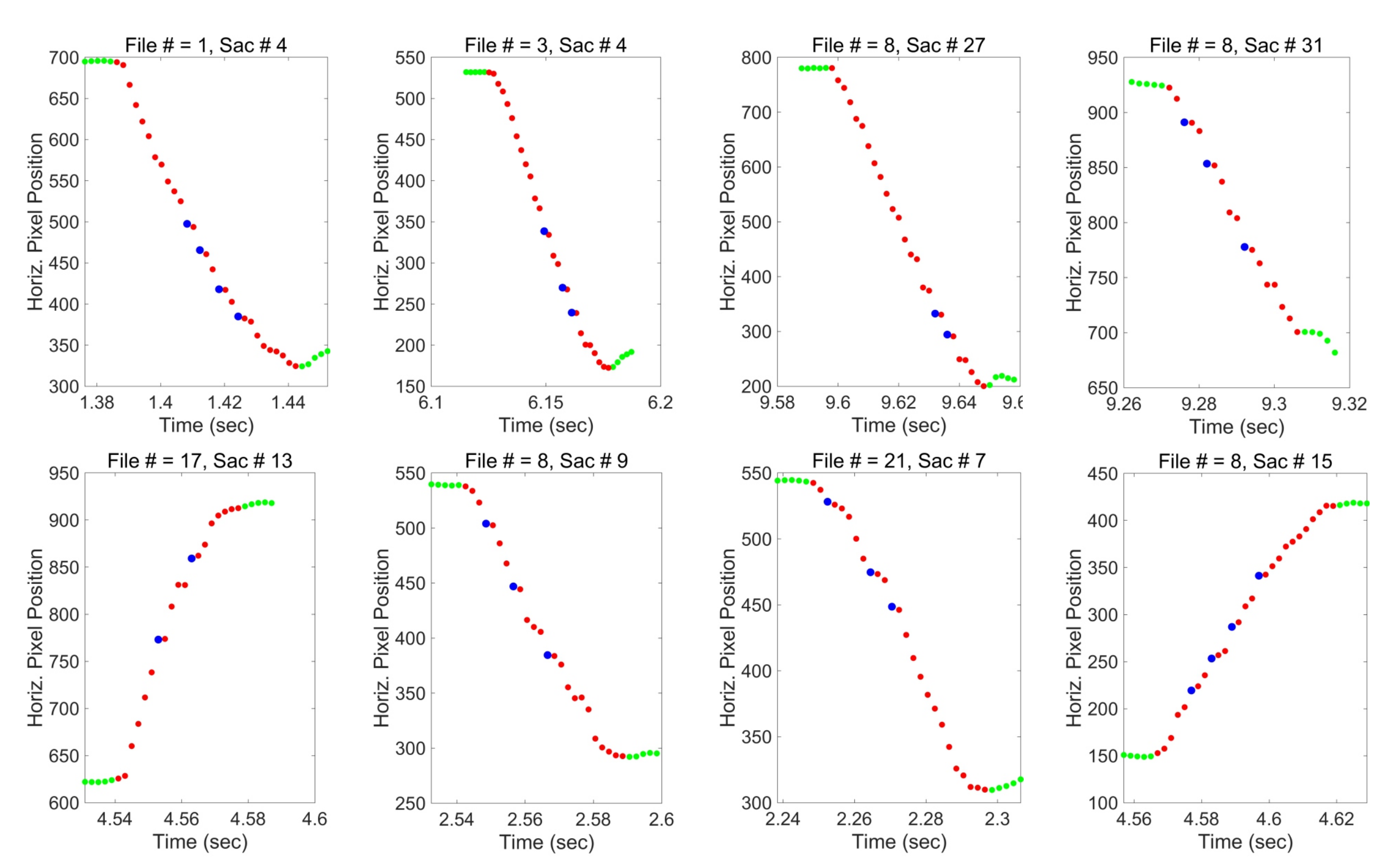}}
\caption{Eight examples, of 8 different saccades from 8 different files from the \textit{Lund2013}. File 1 = TH34-img-Europe-labelled-MN, File 2 = TH34-img-vy-labelled-MN, File 3 = TL20-img-konijntjes-labelled-MN, File 4 = TL28-img-konijntjes-labelled-MN, File 5 = UH27-img-vy-labelled-MN, File 6 = UH29-img-Europe-labelled-MN, File 7 = UH33-img-vy-labelled-MN, File 8 = UL43-img-Rome-labelled-MN.  The title for each plot contains the file number and the saccade number.  The hand scoring contained in the dataset was used to determine where saccades occur (shown in red). Non-Saccade data are shown in green, and the blue dots indicate the occurrence of the unusual discontinuities in this data (see text for explanation). These plots all employed the hand classification performed by rater``MN''. }
\label{fig:2}
\end{figure*}

\begin{figure}[htbp]
\includegraphics[width=0.5\textwidth]{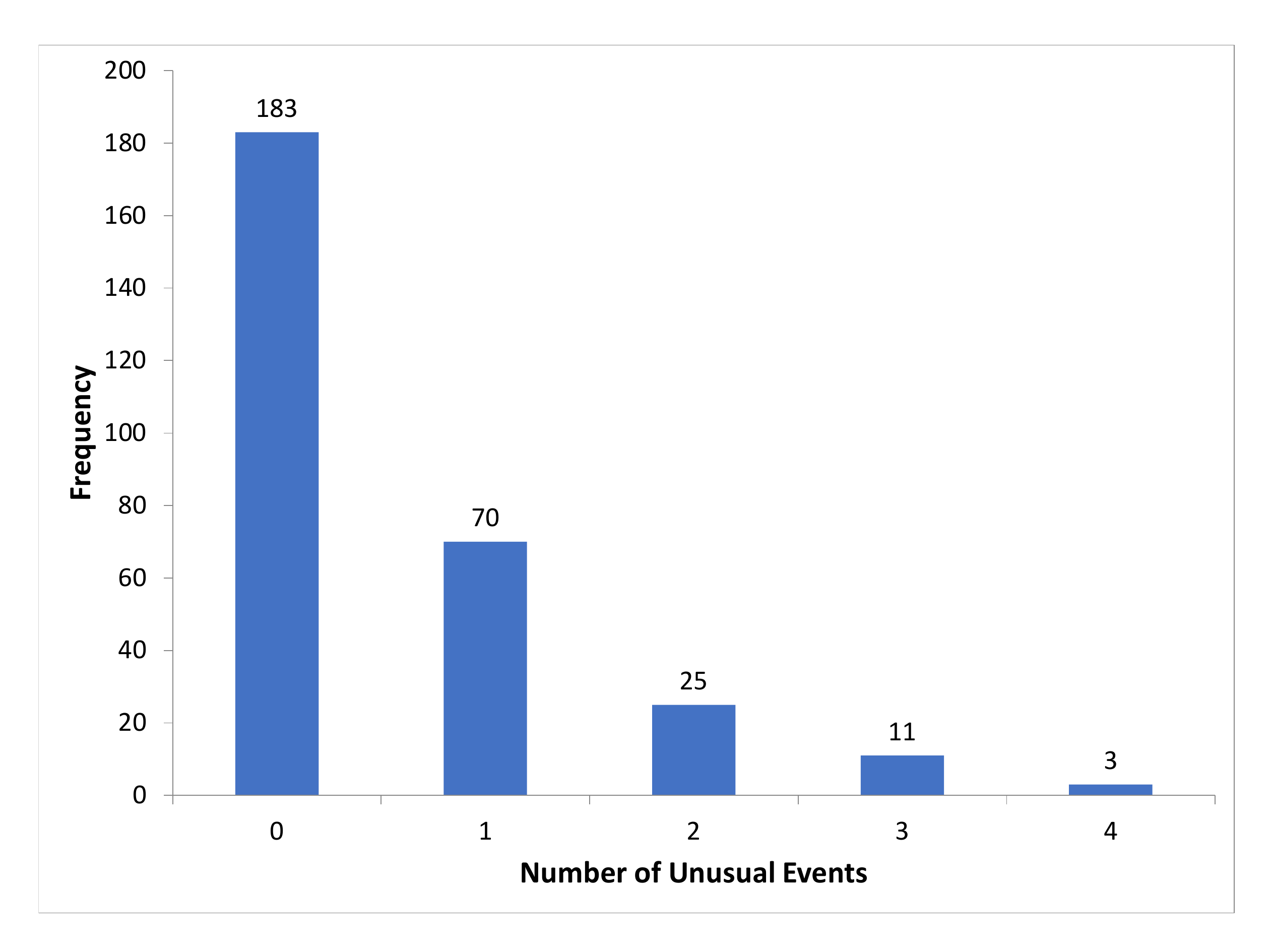}
\caption{Frequency histogram of number of discontinuous events per saccade.}
\label{WierdEventHistogram}  
\end{figure}

These abnormalities are easily spotted when viewing the signals. I certainly think it is highly relevant for the reader to be informed about these unusual saccade artifacts, and their presence certainly should have been reported.  Given all of these issues with the Lund2013, I would not recommend its use, without correction, to develop and test a machine learning algorithm designed to classify eye-movements generally.  


These kinds of discontinuities would obviously lead to an mis-estimation of saccade velocity, which the MNH very much depends on, as does the algorithm of \cite{Nystrom}.  It is not possible to accurately assess the classification accuracy of the MNH on the \textit{Lund2013} until this problem is solved, assuming it can be solved. \interfootnotelinepenalty=10000\footnote{There are classification issues with respect to the GazeCom and humanFixationEvaluation datasets that are also evaluated in the gazeNet paper.  Some of the issues with GazeCom are outlined in \cite{Shagen}, (see section 3.1.4, with 6 illustrations of  misclassified saccades).  This evaluation will be amplified in a future manuscript. For our analysis of the approach gazeNet applies to the humanFixationEvaluation dataset, see "GazeNetCritiqueAddendum.pdf" at  https://digital.library.txstate.edu/handle/10877/8649}

Although the gazeNet paper was developed using eye movement data that had several unusual properties, it is a general method that could, in theory, be applied to any similar eye-movement dataset.

\section{Error Number 2}
\label{ErrorNumber2}
On page 5 of the gazeNet paper, the authors state:  

\begin{adjustwidth}{1cm}{}``...gazeNet was evaluated together with 4 other event detection algorithms on the Lund2013-image-test dataset, genSet and 2 other eye movement datasets.''
\end{adjustwidth}

\noindent On page 17, the authors state:

\begin{adjustwidth}{1cm}{}``The main exception is in the performance of MNH in detecting fixations: at the event-level it outperforms IRF (but not gazeNet) in the lund2013-image-test...''
\end{adjustwidth}

The data on this comparison is prominently displayed in Table 8.

This makes it clear that the MNH and other algorithms were compared to gazeNet on the Lund2013 dataset.

It should be noted that machine learning algorithms are trained to minimize classification error for the provided ground-truth labels. If a great many labeled saccades in the training set have the discontinuities illustrated above, then the network is expected to detect such unconventional events during inference.  As MNH was not designed to identify discontinuous saccades, there is no question that the gazeNet algorithm has an unfair advantage when classifying the Lund2013 dataset, where such discontinuous trajectories are labeled as standard saccades.  But the problem is even more extensive than this.  In addition to saccade discontinuities, the machine learning approach can “memorize” dataset specific patterns (saccade amplitude, frequency of saccades, general motion patterns, etc ...).  Any model comparison is unfair if only 1 algorithm was trained on the data used to compare methods and the other methods were not trained on this dataset.  Nowhere in the gazeNet paper is this inherent unfairness acknowledged.
\footnote{In the paper introducing the MNH \cite{MNH}, 3 other algorithms were compared to the MNH on EyeLink-1000 data collected at 1000 Hz. Two were developed or designed for such EyeLink data.  The third algorithm \cite{Nystrom} was developed on an SMI HiSpeed device. In this case, I received personal communication from Dr. Nyström indicating that that there should be no problem as long as the sampling rate parameter was modified.  Also, Dr. Holmqvist, who self-identified as one of the reviewers, raised no issues about the appropriateness of applying their algorithm to EyeLink data.}

\section{Error Number 3}
\label{ErrorNumber3}
The third error has to do with the calculation of the event-level agreement analysis in the gazeNet paper.\footnote{I received substantial initial assistance related to this error from a graduate student (Vladyslav Prokopenko). He is my colleague in Dr. Komogortsev's laboratory at Texas State University.   Before I got involved, he had looked carefully at the event-level analysis code and told me that he was confident that it did not do what the gazeNet authors intended, but he was not able to describe the exact nature of the error, nor was he able to point to the lines in the code which caused the error. Mr. Prokopenko also provided four Python code assignment statements that corrected the error in the original code.}  As part of that paper, the authors presented a new method to assess event-level agreement between 2 classifiers.\footnote{The Python code which performs the event-level agreement analysis for the gazeNet paper is available at https://github.com/r-zemblys/ETeval. } One classifier was referred to as “ground truth” (“gt”), and in the gazeNet paper, was a human rater. The other classifier was referred to as an algorithm which makes “predictions” (“pr”).  The final result was a Cohen's kappa for each comparison.  

For ground truth (“gt”) I also used the \textit{Lund2013}, described above.  Specifically, the 34 recordings when subjects were looking at images are the focus of this report as was the case for the gazeNet paper. Like the gazeNet paper, I treated all of the \textit{Lund2013} as if it was collected at the nominal sampling rate of 500Hz.  Issues of data quality are not relevant to the point I am trying to make in this section.  The names of the original datasets and the names I used are listed in the Appendix.  Eye-movement data were classified into fixation, saccades, post-saccadic oscillations, smooth pursuit, blinks and undefined.

For my comparison, “predictive” algorithm, I employed the MNH \cite{MNH}.  As the creator of the MNH, it was easy to get the classifications.  I reiterate that data quality is not an issue for the point I want to make regarding event-level agreement. I am not trying to accurately estimate the classification accuracy of the MNH on the \textit{Lund2013}.  The MNH classifies data into fixations, saccades, post-saccadic oscillations, artifact, small RIONEPS noise \cite{MNH,RIONEPS} and large RIONEPS noise.  All sections of each dataset that were not labelled as fixation, saccade or PSO were removed from both the gt scored and the pr scored data.  Any dataset that contained less than 1000 remaining samples was not analyzed further.  This resulted in 28 datasets\footnote{Datasets are at https://doi.org/10.18738/T8/LKSTW4}.

The description of the algorithm in the gazeNet paper does not include pseudocode, nor does it precisely describe the actual steps in the analysis. Also, this description fails to introduce the concept of true negatives, which as we will see below, are critical to consider.  Therefore, I will take some time to introduce this method using my own figures and tables.  The input to the analysis is a two-column dataset, with one column listing the original classes (fixations, saccades and PSOs), on a sample by sample basis, from the gt method (\textit{Lund2013} hand scoring), and one column listing the original classes from the pr (MNH) algorithm.  Fixations were coded as 1, saccades as 2 and PSOs as 3.  Herein, I will be focused exclusively on obtaining kappas for fixation events, saccade events and PSO events.  For illustrative purposes,  I will be focused on the event-related kappa for PSO events for my 12th dataset (“Set\_12.csv”). To produce an evaluation of say, PSO events, all events that are not PSOs are combined into non-PSO events (Appendix II, Figure 1)

The next step was to match gt binary events to pr binary events.  This was done by determining the degree of overlap between events.  The amount of overlap between each gt event and each pr event was calculated.  Then each overlap was considered for matching from the longest to shortest and events were matched based on the overlap, if, and only if, both events were not already matched to other events.  A single pr event can only be matched to a single gt event.  

The data format for input to the kappa function used in this code was two vectors, one representing the class (1 or 0) of the gt event, and the other representing the class (1 or 0) of the pr event.  Unmatched gt events and unmatched pr events are also given similar codes.  The final codes for the PSO evaluation for set number 12 are shown in Table 1.  Gt events that are matched to pr events are labelled as ``Matched''.  Although the codes for these matched events are typically either 1-1 or 0-0, this was not always the case.  In the gazeNet analysis, two events can be matched, but one could be a PSO and the other could be a “not-PSO”.  For example, event \# 14 in this list (Table 1) shows a match, where the gt event was a PSO and the pr event was a not-PSO.   Also, there are unmatched gt events and unmatched pr events.  Although the two vectors that are labelled in Table 1 as gt\_label and pr\_label are all that are required to compute kappa using the particular function employed in the original Python code, statisticians typically convert such information into a contingency table like that shown in Table 2. Once the data are in this form, Cohen's kappa can be easily calculated.

\renewcommand*{\thefootnote}{\fnsymbol{footnote}}
\begin{table*}[htbp]
\center
\caption{Codes for Events to be Input to the Kappa Function. PSO Analysis, Set\_12.csv}
\begin{minipage}{16cm}
\begin{tabular}{|c|c|c|c|c|c|c|c|c|}
\hline
& & & & \textbf{Original} & \textbf{Corrected} & \textbf{Corrected} & \textbf{Correct}\\
\textbf{Count}  & \textbf{Status}  & \textbf{gt\_label} & \textbf{pr\_label} & \textbf{classification\footnote{TN = True Negative, FP = False Positive, FN = False Negative}} & \textbf{gt label} & \textbf{pr label} & \textbf{Classification} \\
\hline
1 & Matched & 0 & 0 & & & &  \\
\hline
2 & Matched & 1 & 1 & & & & \\
\hline
3 & Matched & 0 & 0 & & & &  \\
\hline
4 & Matched & 1 & 1 & & & &  \\
\hline
5 & Matched & 0 & 0 & & & &  \\
\hline
6 & Matched & 1 & 1 & & & &  \\
\hline
7 & Matched & 0 & 0 & & & &  \\
\hline
8 & Matched & 1 & 1 & & & &  \\
\hline
9 & Matched & 0 & 0 & & & &  \\
\hline
10 & Matched & 1 & 1 & & & & \\
\hline
11 & Matched & 0 & 0 & & & &  \\
\hline
12 & Matched & 1 & 1 & & & &  \\
\hline
13 & Matched & 0 & 0 & & & &  \\
\hline
14 & Matched & 1 & 0 & & & &  \\
\hline
15 & Matched & 0 & 0 & & & &  \\
\hline
16 & Matched & 1 & 1 & & & &  \\
\hline
17 & Matched & 0 & 0 & & & &  \\
\hline
18 & UnMatched\_GT & 0 & \cellcolor{blue!45}0 & \cellcolor{red!40}TN & 0  & 1  & FP  \\
\hline
19 & UnMatched\_GT & 1 & \cellcolor{blue!45}0 & FN & & & FN  \\
\hline
20 & UnMatched\_GT & 0 & \cellcolor{blue!45}0 & \cellcolor{red!40}TN & 0  & 1  & FP  \\
\hline
21 & UnMatched\_GT & 1 & \cellcolor{blue!45}0 & FN & & & FN  \\
\hline
22 & UnMatched\_GT & 1 & \cellcolor{blue!45}0 & FN & & & FN \\
\hline
23 & UnMatched\_GT & 0 & \cellcolor{blue!45}0 & \cellcolor{red!40}TN & 0  & 1  & FP \\
\hline
24 & UnMatched\_GT & 0 & \cellcolor{blue!45}0 & \cellcolor{red!40}TN & 0  & 1  & FP \\
\hline
25 & UnMatched\_GT & 1 & \cellcolor{blue!45}0 & FN & & & FN \\
\hline
26 & UnMatched\_PR & \cellcolor{green!45}0 & 1 & FP & & & FP  \\
\hline
27 & UnMatched\_PR & \cellcolor{green!45}0 & 0 & \cellcolor{red!40}TN & 1  & 0  & FN \\
\hline
28 & UnMatched\_PR & \cellcolor{green!45}0 & 1 & FP & & & FP  \\
\hline
29 & UnMatched\_PR & \cellcolor{green!45}0 & 1 & FP & & & FP \\
\hline
\end{tabular}
\end{minipage}
\end{table*}
\renewcommand*{\thefootnote}{\arabic{footnote}}

\begin{table}[htbp]
\center
\caption{Contingency Table}
\begin{tabular}{|c|c|c|}
\hline
& \textbf{PR=1} & \textbf{PR=0}\\
\hline
\textbf{GT=1} & TP(1,1) & FN(1,0) \\
\hline
\textbf{GT=0} & FP(0,1) & TN(0,0) \\
\hline
\end{tabular}
\end{table}

To calculate kappa for this binary form of the data, all events in each input file need to be classified as either True Positive (TP), True Negative (TN), False Positive (FP) or False Negative (FN). 

The results of the matching process and code labelling is illustrated in Table 1 (see also Appendix II Figure 2).  In Table 1, note that some unmatched gt events are labelled 0-0.  In terms of the contingency table presented in Table 2, these would be interpreted as true negatives.  But all unmatched events must be considered as errors, and coded either 1-0 or 0-1.  True-negatives (0-0) are not errors.  I quote from the gazeNet paper:
\\
\begin{adjustwidth}{1cm}{}
``The remaining unmatched events are then labeled as false positives or false negatives, depending on whether they, respectively, occur in the ground truth or the algorithm event stream.'' (page 6, right column, 2nd line)\footnote{There are 4 potential types of mismatches, unmatched gt-true events, unmatched gt-false events, unmatched pr-true events and unmatched pr-false events.  Unfortunately, the gazeNet authors do not specify exactly how each of these should be coded.\\} 
\end{adjustwidth}

\noindent
\\True negatives should only occur when non-PSO gt events are being correctly matched to non-PSO pr events. True negatives are an indication of good performance.  That unmatched events are being labelled as true negatives is the indication that the original code is not performing correctly.  If one looks at Table 1, one will see that all unmatched gt events are have pr codes of 0 (blue highlight).  You will also see that for all unmatched pr events, the corresponding gt code was 0 (green highlight).  Therefore, although all of the unmatched events are errors and should be coded as either 1-0 or 0-1 (false negatives or false positives), many of them (5 of 12 or 42\%) are actually coded as true negatives (red highlight).  Treating unmatched events as true negatives has the effect of increasing kappa toward 1.0, and therefore must be an error.
There are two lines in the code that are responsible for the error (see Table 3).

\begin{table*}[htbp]
\center
\caption{Python Assignment Statements}
\begin{tabular}{|l|c|}
\hline
\textbf{Original code snippets:}                            &  \\
\hline
evt\_pr = pd.concat((evt\_pr, pd.DataFrame(np.zeros(len(set\_gt)))))                            & (1) \\
\hline
evt\_gt = pd.concat((evt\_gt, pd.DataFrame(np.zeros(len(set\_pr)))))                            & (2) \\
\hline
\textbf{Corrected code snippets:}                                                                 &    \\
\hline
inverted\_gt = ($\sim$\_etdata\_gt.evt.loc{[}set\_gt, 'evt'{]}.values.astype(bool)).astype(int) & (3) \\
\hline
inverted\_pr = ($\sim$\_etdata\_pr.evt.loc{[}set\_pr, 'evt'{]}.values.astype(bool)).astype(int) & (4) \\
\hline
evt\_pr = pd.concat((evt\_pr, pd.DataFrame(inverted\_gt)))                                      & (5) \\
\hline
evt\_gt = pd.concat((evt\_gt, pd.DataFrame(inverted\_pr)))                                      & (6) \\
\hline
\end{tabular}
\end{table*}

Prior to the first line of code (1), evt\_pr is a list of the codes (0 or 1) for matched events. The variable “set\_gt” contains a list of all the unmatched gt events.  The part of assignment statement (1) that is:
\begin{center}$pd.DataFrame(np.zeros(len(set\_gt)))$\end{center}

\noindent
is appending a series of 0 codes to the evt\_pr list that is the same length as set\_gt.  
Appending a list of 0 values is simply not correct.  I propose that what is needed is to append a list of codes that are opposite of the paired gt codes.  So, if an event is unmatched, it must be coded either 0-1 or 1-0.  I have prepared a corrected version of the event-level agreement Python code
\footnote{See
https://digital.library.txstate.edu/handle/10877/8649, where you can download ImprovedEventLevelKappa.zip.  This code reproduces the original erroneous kappas and the new, corrected kappas. It does not depend on the original ETeval code.  MATLAB code is also available to graphically illustrate the event-level agreement results, as in Appendix II, Figures 1 and 2.}.
In assignment statements (3) and (4) (Table 3), the codes that will be appended for unmatched events are created as inverted\_gt and inverted\_pr.  If the unmatched gt event has a code of 0, the pr code for the same event will be 1.  If the unmatched pr event has a code of 0, the gt code for the same event will be 1. The original code overestimates the number of true negatives and underestimates the number of false negatives and false positives.  Therefore, the original code produces kappa estimates that are too high.  In Fig. 4, I present distributions of incorrect Kappa\_gN (kappa\_gazeNet) values as well as the corrected Kappa\_Cor values for fixations, saccades and PSOs.

\begin{figure*}[htbp]
\center
\frame{\includegraphics[width=1.0\textwidth]{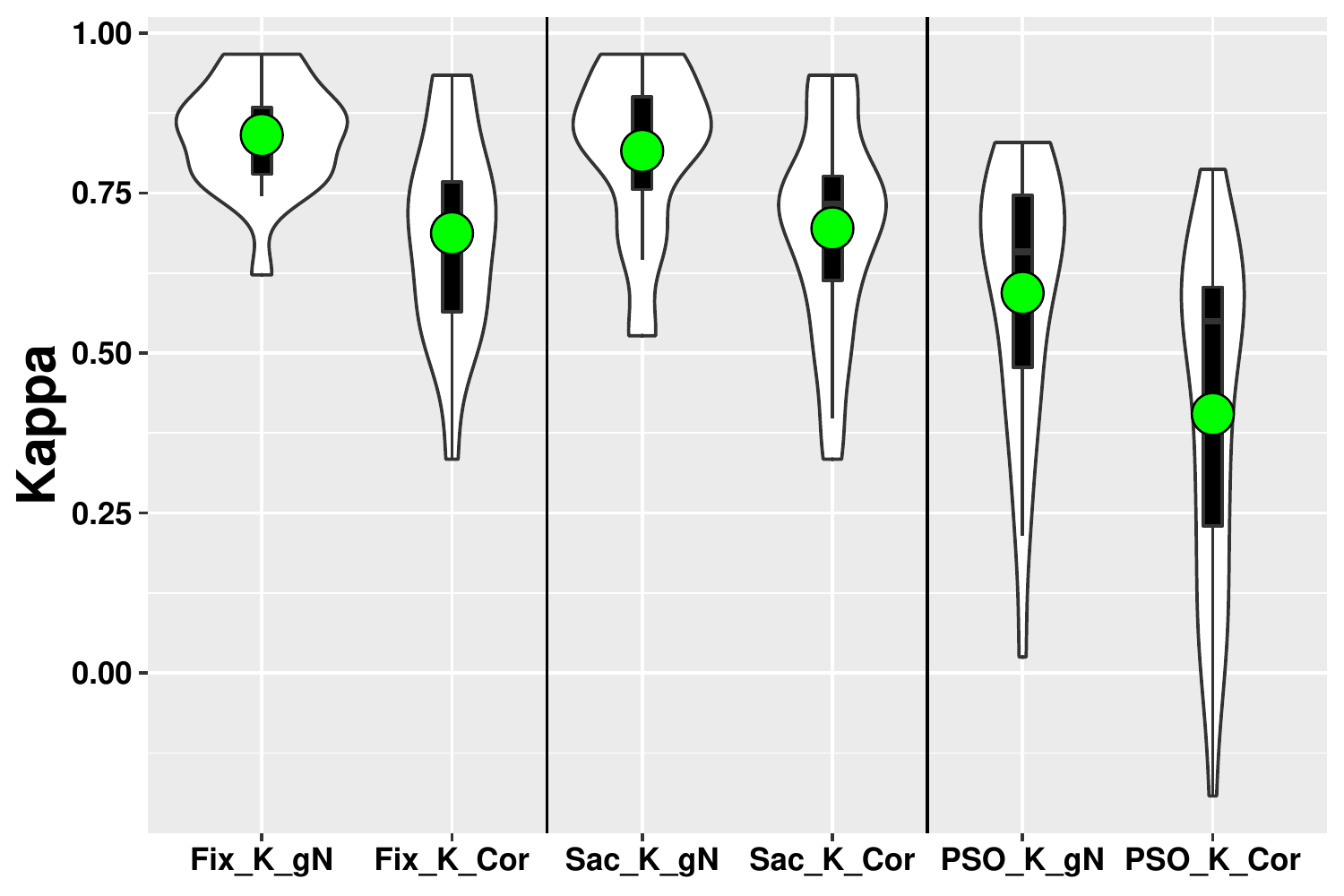}}
\caption{Violin plots of Cohen Kappa values before correction (“gN”, for gazeNet) and after correction (“Cor”, for correct). Mean values are shown in green.  The interquartile range is shown as a black vertical bar. For fixation, there was a decrease of 18.27\% after correction in mean kappa (t = 9.18, df = 19, p-value = 0.00000002, paired t-test, 2-tailed). For saccades, there was a decrease of 16.20\% after correction (t = 6.55, df = 22, p-value =0.000001).  For PSOs, there was a decrease of 33.78\% (t = 8.61, df = 26, p-value =0.000000004). Cases where kappa\_gN was equal to 1.0 were not included in this analysis (Fixation N = 8, Saccade N = 5, PSO N = 1), since there was nothing to correct in these cases. In three cases, all involving PSOs, the percent decrease was greater than 70\% (0.42 to 0.1, 0.35 to 0.08 and 0.42 to 0.03).  If cases in which kappa\_gN = 1.0 are included, the mean percent decrease across all events was 21.1\%, but in 8 cases the percent decrease was greater than 50\%.}
\label{fig:6}
\end{figure*}

The gazeNet authors make a strong case for using kappa over an F1-score for sample-level agreement statistics.  With regard to the event-level they state:\\

\begin{adjustwidth}{1cm}{}
``Note that when evaluating the detector’s performance at the event level instead of the sample level, the distribution of data across the classes is less unbalanced. Nevertheless, Cohen’s kappa is also a robust measure to use in this case, and will thus be used throughout the paper.''
\end{adjustwidth}
\noindent
\\There is an alternative way to describe the situation.  With binary classifications, every true event is followed by a false event. Thus, the distribution of classes is either perfectly balanced or only trivially unbalanced.  An F1-score would be appropriate. Other researchers use the F1-score to describe their event-level agreement results \cite{HoogeF1,Startsev}. There is no concept of a true negative in the context of the F1-score, and there is no ambiguity regarding the classification of false negatives (Misses) and false positives (False Alarms).  The gazeNet authors could have preferred one matching criteria over another, (e.g., “largest overlap” versus “overlapping test-events occurring earliest in time”), and still analyzed the performance using an F1-score. 

Also, in the section in the gazeNet paper describing this event-level evaluation there were two additional errors.  The statement: \\

\begin{adjustwidth}{1cm}{}
``Hoppe and Bulling  \cite{HopBul} used a majority vote approach and calculated the F1-score.''\\
\end{adjustwidth}

\noindent
is false.  Hoppe and Bulling \cite{HopBul} do not express their event-level agreement results in terms of F1-scores.  

Also, in this section, the gazeNet authors make the following statement:\\

\begin{adjustwidth}{1cm}{}
``Hooge et al. (2017) developed another event level F1-score and used it to compare human coders for assessing fixation coding performance. Hooge et al. looked for overlapping test-events occurring earliest in time and labeled them as hits. The remaining unmatched ground truth fixations were labeled as misses, while the remaining unmatched test fixations were labeled as false alarms. This works when there is a single event class, but when there is more than one event type to match, the Hooge et al. approach does not work well.'' (page 6, left column, near top)
\\
\end{adjustwidth}

\noindent
This notion does not make sense to me. The Hooge et al approach \cite{HoogeF1} is designed to compute an F1-score, and an F1-score is an average of precision and recall.  Precision and recall are only defined in the 2-class case.  They are not defined in the multi-class case.

\section{A Note on Nomenclature}
\label{Nomenclature} In the machine learning literature, including the gazeNet paper, criteria which emerge from human observation, intelligence, insight and intuition are pejoratively refered to as ``hand-crafted".  The hand has nothing to do with such criteria. Jewelry is handcrafted.  Bronze sculpture is handcrafted.  I propose that such criteria be properly referred to as criteria which emerge from human observation, intelligence, insight and intuition ($OI^3$).  Alternatively, criteria emerging from traditional approaches might be referred to as ``mindful'' whereas machine learning approaches might be referred to as ``mindless''.

    An earlier version of this manuscript is posted on arXiv \cite{arXiv}

\section{acknowledgements}
I would like to express my gratitude to Mr. Vladyslav Prokopenko for all of his assistance with this work.  Another graduate student colleague, Mr. Dmytro Katrychuk, made changes to my GitHub repository that facilitates setting up the working environment to use the code. He also significantly improved the Python script to increase its comprehensibility and to remove the dependency on the original ETeval code base.  I very much appreciate his assistance.

\section*{Conflict of interest}
I have no conflict of interest.

\onecolumn

\section{Appendix I}

\textit{Lund2013} dataset names.

\begin{table*}[hbt!]
\center
\begin{tabular}{|l|l|l|l|}
\hline
Original File Names                     & Renamed As  & Original File Names                            & Renamed As  \\
\hline
TL28\_img\_konijntjes\_labelled\_MN.mat & Set\_01.csv & UL23\_img\_Europe\_labelled\_RA.mat            & Set\_18.csv \\
\hline
TH34\_img\_Europe\_labelled\_MN.mat     & Set\_02.csv & UL31\_img\_konijntjes\_labelled\_RA.mat        & Set\_19.csv \\
\hline
TH34\_img\_vy\_labelled\_MN.mat         & Set\_03.csv & UL39\_img\_konijntjes\_labelled\_RA.mat        & Set\_20.csv \\
\hline
TL20\_img\_konijntjes\_labelled\_MN.mat & Set\_04.csv & UL43\_img\_Rome\_labelled\_RA.mat              & Set\_21.csv \\
\hline
UH21\_img\_Rome\_labelled\_MN.mat       & Set\_05.csv & \cellcolor{blue!45}UL47\_img\_konijntjes\_labelled\_RA.mat        & Set\_22.csv \\
\hline
UH33\_img\_vy\_labelled\_MN.mat         & Set\_06.csv & TH38\_img\_Europe\_labelled\_RA.mat            & Set\_23.csv \\
\hline
UL23\_img\_Europe\_labelled\_MN.mat     & Set\_07.csv & TH46\_img\_Rome\_labelled\_RA.mat              & Set\_24.csv \\
\hline
UL31\_img\_konijntjes\_labelled\_MN.mat & Set\_08.csv & TH50\_img\_vy\_labelled\_RA.mat                & Set\_25.csv \\
\hline
UL39\_img\_konijntjes\_labelled\_MN.mat & Set\_09.csv & TL44\_img\_konijntjes\_labelled\_RA.mat        & Set\_26.csv \\
\hline
UL43\_img\_Rome\_labelled\_MN.mat       & Set\_10.csv & \cellcolor{blue!45}TL48\_img\_Europe\_labelled\_RA.mat            & Set\_27.csv \\
\hline
\cellcolor{blue!45}UL47\_img\_konijntjes\_labelled\_MN.mat & Set\_11.csv & \cellcolor{blue!45}TL48\_img\_Rome\_labelled\_RA.mat              & Set\_28.csv \\
\hline
TL28\_img\_konijntjes\_labelled\_RA.mat & Set\_12.csv & UH27\_img\_vy\_labelled\_RA.mat                & Set\_29.csv \\
\hline
TH34\_img\_Europe\_labelled\_RA.mat     & Set\_13.csv & UH29\_img\_Europe\_labelled\_RA.mat            & Set\_30.csv \\
\hline
TH34\_img\_vy\_labelled\_RA.mat         & Set\_14.csv & \cellcolor{blue!45}UH47\_img\_Europe\_labelled\_RA.mat            & Set\_31.csv \\
\hline
TL20\_img\_konijntjes\_labelled\_RA.mat & Set\_15.csv & UH27\_img\_vy\_labelled\_MN.mat                & Set\_32.csv \\
\hline
UH21\_img\_Rome\_labelled\_RA.mat       & Set\_16.csv & UH29\_img\_Europe\_labelled\_MN\_Corrected.mat & Set\_33.csv \\
\hline
UH33\_img\_vy\_labelled\_RA.mat         & Set\_17.csv & \cellcolor{blue!45}UH47\_img\_Europe\_labelled\_MN.mat            & Set\_34.csv\\
\hline
\end{tabular}
\end{table*}

\section{Appendix II}

\textit{Additional Figures} 

\begin{figure*}[htbp]
\center
\frame{\includegraphics[width=1.0\textwidth]{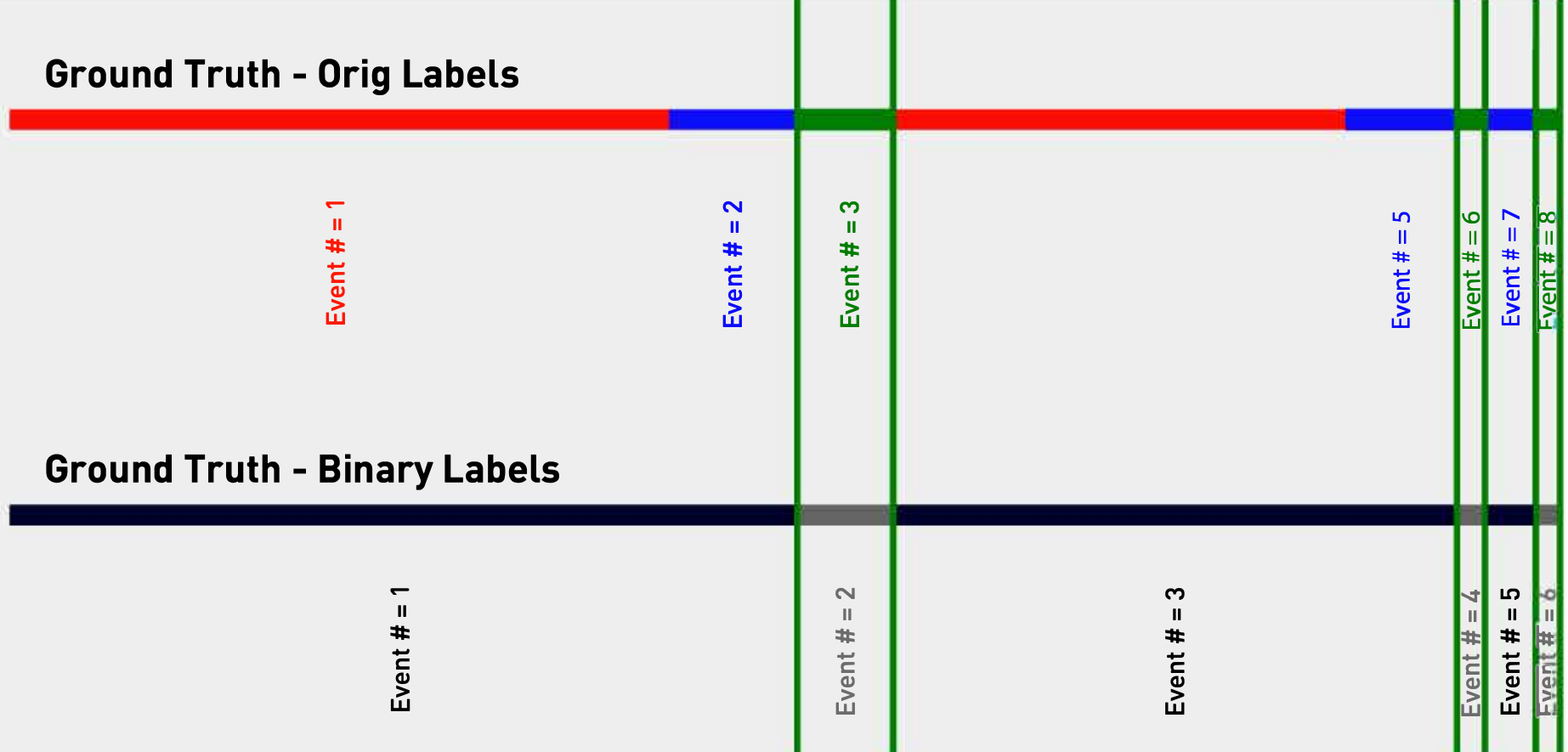}}
\caption{Appendix II. Fig. 1. Fig. 1 Original 3-level classification reduced to a binary classification: PSO vs Non-PSO. In the original data from gt, I have
fixations (red), saccades (blue) and PSOs (green). After conversion to binary labels, Non-PSOs are shown in
black and PSO events are shown in grey. The first event, a fixation, and the second event, a saccade, are grouped together as a
single non-PSO event. The PSO labelled in the original series as event number 3 becomes the second binary event, PSO=true.
A similar conversion to binary labels was made to the pr events.}
\label{AppendixFigure1}
\end{figure*}

\begin{figure*}[htbp]
\center
\frame{\includegraphics[width=1.0\textwidth]{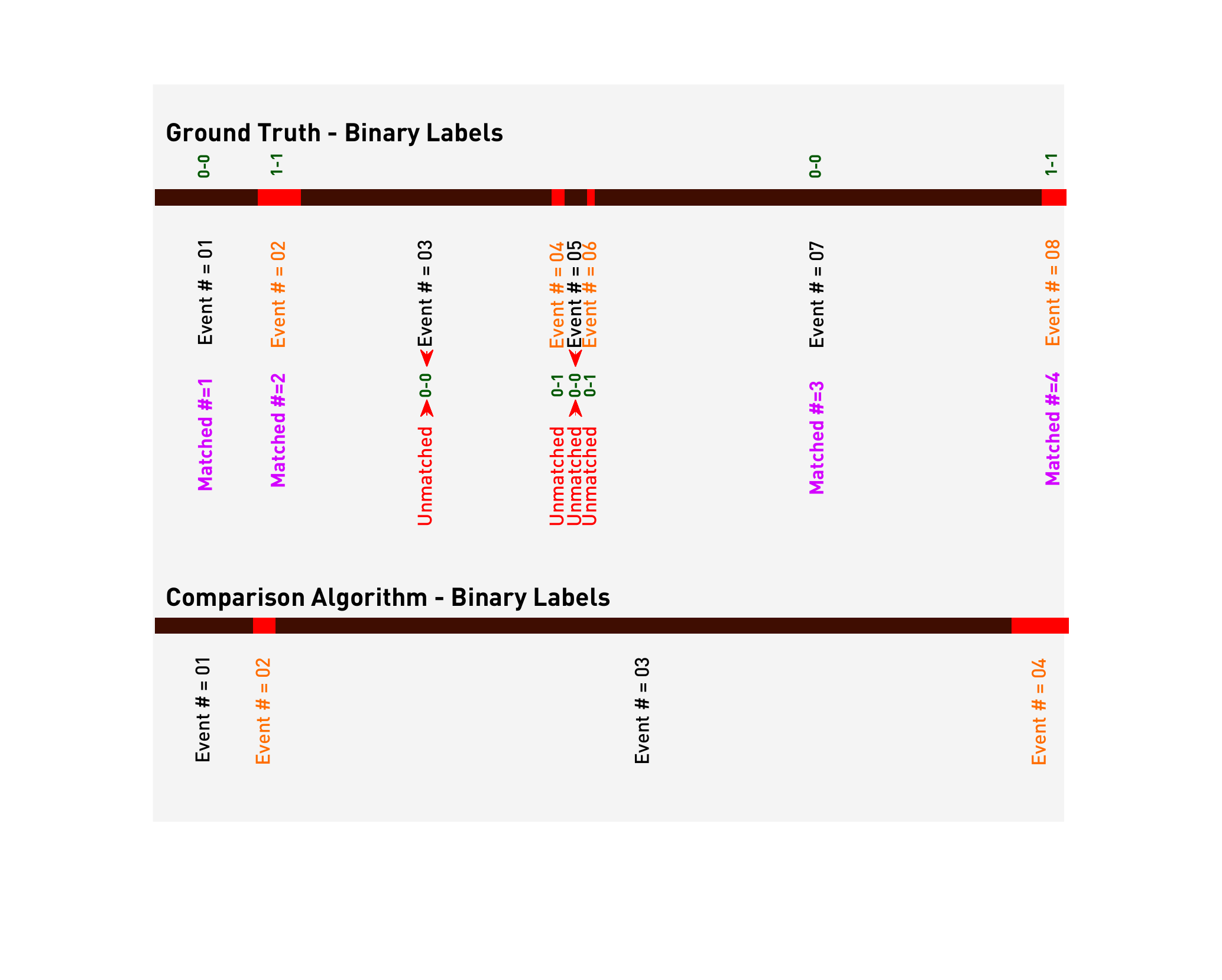}}
\caption{Appendix II. Fig. 2. Binary event matching and relevant codes. The first gt event (\#01) was a not-PSO and it was matched to the first pr event.
The second gt event was a PSO event, and it was matched to the second pr event. GT events \#03, \#04, \#05 and \#06 are not
matched but gt event \#07 was a notPSO and was matched to pr event \#03.  Gt event \#08 was matched to pr event \#04. The green codes (``0-0'', ``1-1'' and ``0-1'') refer to the final labels that are to be input to the Python kappa function. For matched events, the codes appear above the gt line, and for unmatched events, the codes appear above the word ``unmatched''. The lower code refers to the code assigned to pr, and the upper code refers to the code assigned to gt. Unmatched pr events do occur in this data but are not shown in this figure. The red arrows indicate unmatched events that are coded as true negatives (``0-0''). For illustrative purposes, the first event in both the gt and pr were truncated.}
\label{AppendixFigure2}
\end{figure*}

\end{document}